\documentclass[useAMS,usenatbib,usegraphicx]{mn2e}

\usepackage{amssymb}
\begin{document}

\title[Time-Dependent Lyman $\alpha$ Transfer]
{Time-Dependent Behavior of Lyman $\alpha$ Photon
 Transfer in High Redshift Optically Thick Medium}

\author[Xu, Wu and Fang]
{Wen Xu$^{1,2}$, Xiang-Ping Wu$^1$, and Li-Zhi Fang$^2$ \\
$^{1}${{National Astronomical Observatories, Chinese
Academy of Sciences, Beijing 100012, China}}\\
$^{2}${{Department of Physics, University of Arizona,
Tucson, AZ 85721, USA}}\\
}
\maketitle

\begin{abstract}

With Monte Carlo simulation method, we investigate the time dependent behavior of Ly$\alpha$
photon transfer in optically thick medium of the concordance $\Lambda$CDM universe.
At high redshift, the Ly$\alpha$ photon escaping from optically thick medium has a time scale
as long as the age of the luminous object, or even comparable to the age of the universe.
In this case, time-independent, or stationary solutions of the Ly$\alpha$ photon transfer
with resonant scattering will overlook important features of the escaped Ly$\alpha$ photons in
physical and frequency spaces. More seriously, the expansion of the universe leads to that the
time-independent solutions of the Ly$\alpha$ photon transfer may not exist. We show that 
time-dependent solutions sometimes are essential for understanding the Ly$\alpha$ emission and absorption at 
high redshifts. For Ly$\alpha$ photons from sources at redshift $1+z=10$ and
being surrounded by neutral hydrogen IGM of the $\Lambda$CDM universe,
the escape coefficient is found to be always less, or 
much less than one, regardless of the age or life time of the sources. 
Under such environment, we also find that
even when the Ly$\alpha$ photon luminosity of the sources is stable, the mean surface
brightness is gradually increasing in the first 10$^6$ years, and then decreasing with a power
law of time, but never approaches a stable, time-independent state. That is, all $1+z=10$ sources in
 a neutral Hubble expanding IGM
with Ly$\alpha$ luminosity $L$ have their maximum of mean surface brightness
$\sim 10^{-21}(L/10^{43} {\rm erg/s})$ erg s$^{-1}$ cm$^{-2}$ arcsec$^{-2}$ at the age of
about 10$^6$ years. The time-dependent effects on the red damping wing profile are also addressed.

\end{abstract}


\section{Introduction}

Ly$\alpha$ photons have been widely applied to study the physics of
the universe at redshifts from 2 to 8. The redshifted Ly$\alpha$
photons carry the information of the photon source, the halo
surrounding the source, and the IGM at early universe. The optical
afterglow of gamma-ray burst (GRB) has been modeled as the red wing
of Ly$\alpha$ photon absorption, and used to estimate the column
number density of neutral hydrogen of IGM at high redshift
\citep*{Totani06,Salvaterra09}. The transmitted flux of QSO
absorption spectrum at redshift $z> 5$ consists of complete
absorption troughs separated by spikes (e.g., \citealt{Becker01,
Fan06}). The spikes have been explained as Ly$\alpha$ photons
leaking at low density areas (e.g. \citealt{Liu07,Feng08}), and used
to probe the turbulent behavior of IGM at high redshifts
\citep*{Lu10,Zhu10,Zhu11}. The last but not the least, searching for redshifted
Ly$\alpha$ photons from  star forming galaxies at high redshift  is
believed to be a basic tool to explore the epoch of reionization
(e.g. \citealt{Hayes10,Lehnert10}). Therefore, it is crucial to have a complete
understanding of the radiative transfer of Ly$\alpha$ photons caused by
their resonant scattering with neutral hydrogen atoms.

The radiative transfer of Ly$\alpha$ photons in a medium consisting of
neutral hydrogen atoms has been extensively studied either
analytically or numerically.  Yet, there are very few solutions
on the time-dependent behavior of Ly$\alpha$ photon transfer 
 \citep*{Field58, Rybicki94, Higgins09}.  All other
analytical solutions are time-independent based on the Fokker-Planck
equation \citep*{Harrington73,Neufeld90,Dijkstra06}.
Time-independent solutions are important but can only be used to
describe the ``limiting asymptotic behaviors" of the radiative
transfer \citep*{Adams75,Bonilha79}. They tell us nothing about the 
time scales of the radiative transfer of Ly$\alpha$ photons. These 
time-independent solutions can not describe the Wouthuysen-Field (W-F)
effect \citep*{Wouthuysen52, Field58, Field59}, which is essential for
the 21 cm emission and absorption of neutral hydrogen at high redshift
(e.g. \citealt{Roy09b}). It is because the Fokker-Planck approximation would miss
the detailed balance relation of resonant scattering, which is necessary to 
keep the W-F local thermalization \citep*{Rybicki06}. 

Numerical method based on the Monte Carlo (MC) simulation is also popular in
solving the transfer of resonant photons (e.g. \citealt{Loeb99,
Zheng02,Tasitsiomi06,Verhamme06,Laursen07,
Dijkstra08,Pierleoni09,Xu10}). However, there are very few
works dealing with time-dependent problems. For instance, the
time-scale of the W-F local thermalization is still absent in these
studies.

Time-dependent behavior of the Ly$\alpha$ photon transfer is especially
important to understand observations of Ly$\alpha$ photons at high
redshifts. First, either the life time or the age of photon sources at
high redshift is generally short. Second, the optical depth of the IGM
or the halo cloud around the sources is generally large. If the time
scale of the transfer of Ly$\alpha$ photons is comparable to the life-time 
or the age of the photon source, the ``limiting asymptotic state"
will never be approached. One more problem is caused by the cosmic expansion, 
of which the time scale is short at high redshifts. When the
time scale of the cosmic expansion is comparable to that of the resonant photon
transfer in optically thick medium, time-independent solutions probably
do not exist.

Recently, a state-of-the-art numerical method based on the WENO
scheme has been introduced to solve the integro-differential
equation of the radiative transfer of resonant photons
\citep*{Roy09a, Roy09b, Roy10}. It reveals many interesting features
of the transfer of Ly$\alpha$ photons in an optically thick medium,
which cannot be seen with the time-independent solutions of the 
Fokker-Planck approximation. For
instance, it shows that the time scale of the formation of the W-F
local thermal equilibrium actually is short, only about a few
hundred times  of the resonant scattering. The double peaked frequency 
profile of Ly$\alpha$ photon can not be described by time-independent 
analytical solutions unless the optical depth of $\nu_0$ photons is as large as about
10$^6$. This result directly indicates the need of time-dependent
solutions.

The WENO algorithm of the integro-differential equation of the
radiative transfer is fine, but like other high order scheme with
fixed grid without artificial-viscosity, the computation time is
much more than the Monte Carlo method. Therefore, the WENO method
would be not easy to deal with cases of medium with very high
optical depth. The goal of this paper is two-fold. The first is to
show that the Monte Carlo simulation method can properly match the
results of WENO method on the time-dependent features of moderate
optical depth. Secondly, we study the time-dependent solutions of
Ly$\alpha$ photons escaped from optically thick medium. We will not
work on specific objects, but focus on the general time dependent
features which will affect the observability of the Ly$\alpha$
sources embedded in, or behind optically thick medium.

The paper is organized as follows. Section 2 presents the basic
models we will study. The Monte Carlo simulation method and its
tests for time-dependent problems of Ly$\alpha$ photon transfer in
an optically thick medium will be given in Section 3.
The major results of the time-dependent solutions
of Ly$\alpha$ photon emission in the DLA and IGM models are given in \S 4 and 5,
respectively. Problems of absorption by optically thick medium are presented
in \S 6. Discussions and conclusions are given in \S 7.
All the relevant formulae of the radiative transfer of Ly$\alpha$ photons
and the details of MC simulation are presented in Appendix.

\section{Problem}

We study the time dependent transfer of Ly$\alpha$ photons in two
typical models of neutral hydrogen $HI$ medium. The first one is the
so-called DLA (damped Ly$\alpha$ system) halo model, in which a
source is surrounded by a static spherical halo of physical radius $r_p$,
consisting of homogeneously distributed neutral hydrogen with number
density $n_{\rm HI}$ and temperature $T$. The second one is a source
at redshift $(1+z)=10$ located in homogeneously Hubble expanding
IGM, of which the density and temperature are given by the
parameters of the concordance $\Lambda$CDM universe. We call it IGM model.
The radiative transfer equation of the two models are given in Appendix A.

In order to compare with the WENO solutions, for the DLA halo model
we use dimensionless time and radial coordinate, defined,
respectively, as $\eta=cn_{\rm HI}\sigma_0 t$ and $r=n_{\rm
HI}\sigma_0 r_p$, where $t$ and $r_p$ are the physical variables of
time and radial coordinate, and $\sigma_0/\sqrt{\pi}$ is the cross
section of scattering at the resonant frequency
$\nu_0=2.46\times10^{15}$ s$^{-1}$. Therefore, $\eta$ and $r$ are
the time and length in the units of mean free flight-time and mean
free path of photon $\nu_0$, respectively. The value of $r$ actually
is equal to the optical depth of the spherical halo from $r=0$ to
$r$ at frequency $\nu_0$. For a signal propagating in the radial
direction with the speed of light, we have $r= \eta+ {\rm const}$.

As usual, in frequency space, we use variable $x\equiv (\nu-\nu_0)/\Delta
\nu_D$, where $\Delta \nu_D=\nu_0v_T/c= \nu_0 \sqrt{2k_BT/m_{\rm
H}c^2}=1.06\times 10^{11}(T/10^4)^{1/2}$ Hz is the Doppler
broadening at frequency $\nu_0$ by thermal motion $v_T$ of gas with
temperature $T$. The variable $x$ is then the deviation of frequency
$\nu$ from $\nu_0$ in units of the Doppler broadening.

With the dimensionless variables, the specific number intensity of photons is $I(\eta,
r, x, \mu)$, where $\mu=\cos \theta$ being the direction
relative to the radial vector $r$. Thus, all the solutions of
$I(\eta, r, x, \mu)$ do not refer to a specific density $n_{\rm
HI}$ and size $r_p$ (see Appendix \S A). This helps to see the
common features of the DLA halo model.

The optical depth of a halo or cloud with column density $N_{\rm
HI}$ at frequency $x$ is
\begin{equation}
\tau(x)= N_{\rm HI} \sigma(x) = \tau_0 \phi(x,a)
\end{equation}
where $\sigma(x)$ is the scattering cross section, $\tau_0= N_{\rm
HI}\sigma_0$, and $\phi(x,a)$ is the normalized Voigt profile given
by\footnote{Due to different normalization scheme of the Voigt
function of Eq.(2), our definition of $\sigma_0$ is different from
the one used in some literatures by a factor $\pi^{1/2}$.
Consequently, the expressions for mean flight time, mean free path
and optical depth may be different by a factor of $\pi^{1/2}$.}
\begin{equation}
\phi(x,a)=\frac{a}{\pi^{3/2}}\int^{\infty}_{-\infty}\frac{e^{-y^2}}{(x-y)^2+a^2}dy,
\end{equation}
where the parameter $a$ is the ratio of the natural to the Doppler
broadening. For the Ly$\alpha$ line, $a=4.7\times
10^{-4}(T/10^4)^{-1/2}$. The profile Eq.(2) describes the joint
effect of the Gaussian distribution of the velocity of neutral
hydrogen atom and the Lorentz profile of cross
section in the rest frame of the atom. For an expanding (or
collapsing) halo or turbulent gas cloud, the bulk velocity of the gas
might be larger than the thermal velocity $v_T$. Even in this
case, the Doppler or thermal broadening is still important, as it is
the key factor leading to a local thermal equilibrium of Ly$\alpha$
photons (the W-F effect).

In an optically thick spherical cloud, most time-dependent behaviors
can be described by two time-dependent distribution functions: 1.)
angularly averaged, $r^2$-rescaled, specific  number intensity
$j(\eta,r,x)=(r^2/2)\int_{-1}^{+1}I(\eta,r,x,\mu)d\mu$, and 2.) $r^2$-rescaled number flux
$f(\eta,r,x)=(r^2/2)\int_{-1}^{+1}\mu I(\eta,r,x,\mu)d\mu$, which
describes the photons escaped from a spherical halo of radius $r$.
The equations of $j$ and $f$ are given in Appendix Eqs.(A1) and
(A2).

For the IGM model, we use the physical variable $t$ and
$r_p$ because of the specific cosmological model we adopted.
The temperature of IGM is taken as $T=100$ K and the parameter
$\gamma=1/\tau_{GP}$ [see Eq.(A3)], which describes the Hubble
expanding, is equal to $ 1.4\times 10^{-6}$ in the concordance
$\Lambda$CDM universe.

\section{Method and Test}

\subsection{Time dependent Monte Carlo simulations}

We use Monte Carlo (MC) method to simulate $j(\eta,r,x)$
and $f(\eta,r,x)$ of the previous
section. Most Monte Carlo codes of simulating time-independent
solutions of radiative equation can easily be modified to deal with
time-dependent problems. If the feedback of photon transfer on the
parameters of neutral hydrogen is negligible, the radiative transfer
equations are linear with respect to $I$, and thus to $j$ and $f$.
Thus,  linear superposition of the solutions of the sources
is valid. The time-dependent solutions can simply be given by a
weighted summation over the results of a single flash. The weight
of the summation is proportional to the time-dependent flux of the light
source.

We will employ the same MC algorithm as used in
\citet{Xu10}, of which some of the details are given in Appendix B and C.
The major modification from the earlier methods, such as
\citet{Zheng02}, is to record the time at each collision of the
photon, and to take snapshots of photon distribution in spatial and
frequency space based on these time stamps. Thus, the time-dependent
solutions $j$ and $f$ of an arbitrary source can be given by a
synthesis of the fluxes at different
epochs from a single flash source.

\subsection{Test with $x$-profile of flux}

As a test of the MC method, we calculate the flux $f(\eta, r,x)$ at
the outer boundary $r_{\rm D}=10^2$ of a DLA halo. This result is
shown in Fig. 1, in which the curves are the MC results with
parameter $\eta=500$, 2000 and 3000. They show typical
double-peaked profile.  The curves of $\eta=2000$ and 3000 actually
are the same. The overlapping indicates that $f(\eta, r,x)$ is already at the limiting
asymptotic state, or saturated for time $\eta \geq 2000$. In Fig.
1, the data points are given by the WENO numerical
solutions of \citet*{Roy10}. Therefore, the MC method can match the
time-dependent WENO solutions.

\begin{figure}
\includegraphics[width=8.0cm]{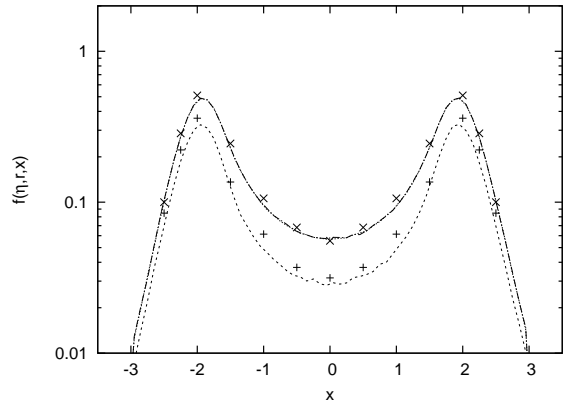}
\caption{The Ly$\alpha$ photon flux $f(\eta, r, x)$ from a DLA halo with radius $r_{\rm
D}=10^2$ and parameter $a=10^{-3}$. The curves are from MC simulations at time
$\eta=500$, 2000 and 3000, respectively from bottom up.
The last two curves (2000 and 3000) are already overlapped with each other.
The symbol points marked by ``+" and ``x" are from the WENO numerical solution
of the radiative transfer equation \citep*{Roy10} at epochs
$\eta=500$, 2000, respectively.} \label{fig:fig1}
\end{figure}

In the WENO method, the solutions of the angularly averaged specific
intensity $j(\eta, r,x)$ and flux $f(\eta, r,x)$ at boundary $r_{\rm
D}$ should satisfy the condition $j(\eta, r_{\rm D},x)=2f(\eta,
r_{\rm D},x)$ (Unno 1955).
This is the result of Eddington approximation and the assumption of no incoming
 photons at the boundary. Our MC simulation
results of $j(\eta, r_{\rm D} , x)$ at time $\eta = 500$ are shown at various radius in Fig. 2.
 At the surface where $r = 100$, we see $j \approx 2f$ at the center frequencies when compared
with Fig.1.  It shows that the MC simulation can
 pass the test of Unno's boundary condition. We also find that $j = 2f$
relation is only valid at the boundary at center frequencies.
Slightly beneath the surface, at radius $r = 99, 98$, we
find that $j \approx 4f$ , and $5f$ , respectively. The enhanced
photon intensity is a result of backward scattered
photons near the boundary.

\begin{figure}
\centering
\begin{center}
\includegraphics[width=8.0cm]{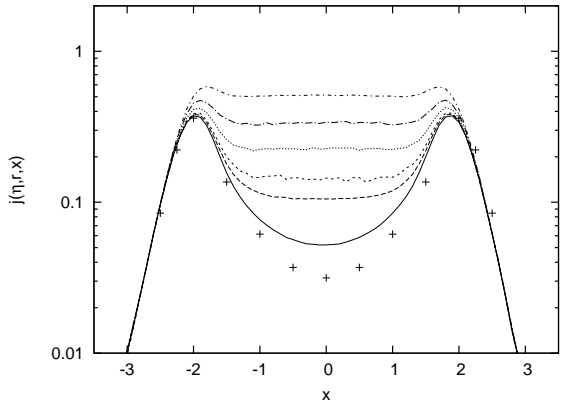}
\end{center}
\caption{The MC simulation results of the angularly averaged specific
intensity $j(\eta, r,x)$ at time
$\eta=500$ for DLA halo model of  parameter $a=0.001$ with size $r_{\rm D}=10^2$. The
curves correspond to $r=$ 100, 99, 98, 95, 90, 80,
respectively from bottom up.  The flux $f$ at time $\eta = 500$ at the boundary $r = 100$
from Fig.1 is marked with ''+" symbols for comparison.
}
\end{figure}

Besides the curve of $r=r_{\rm D}=100$, Fig. 2 also plots $j(\eta,
r,x)$ at time $\eta=500$, but for $r= 80$, 90, 95, 98, 99. We see
that all the curves of $r<100$ are almost flat in the range of
$|x|<2$. It means that the frequency distribution of photons is
thermalized near the resonant frequency $\nu_0$. That is, the
frequency distribution is of Boltzmann $\propto \exp(-\frac{h\Delta \nu_D}{kT} x)$, where
$T$ is the kinetic temperature of neutral hydrogen gas in the halo.
This is the W-F local thermalization  \citep*{Wouthuysen52,
Field58,Field59}. Fig. 2 tells us that the W-F local thermalization
is achieved by resonant scattering even at $r=99$. Yet photons at the
outermost layer $r=r_{\rm D}$ have not yet been thermalized, as the
optical thin layer does not carry enough number of scattering. This
result is similar to the WENO solution \citep*{Roy09b,Roy10}.

\subsection{Test with $|x_{\pm}|$-$\tau_0$ relation}

The second test is on the $|x_{\pm}|$-$\tau_0$ relation, where
$x_{\pm}$ are frequencies of the two peaks of the double-peaked
profile as shown in Fig. 1, and $\tau_0$ is the optical depth
parameter defined in Eq.(1). This relation has been studied by many
time-independent solutions based on the Fokker-Planck equation. The
major conclusion is the so-called $(a\tau_0)^{1/3}$-law, i.e. $x_{\pm} =
\pm A(a \tau_0)^{1/3}$, where $A$ is a constant of order 1
\citep*{Adams72, Adams75, Harrington73, Neufeld90, Dijkstra06}. It
is well known that the $(a\tau)^{1/3}$-law is available only when
the optical depth $\tau_0$ is very large. However, the
$x_{\pm}$-$\tau_0$ relation available for various $\tau_0$ has not
been calculated until very recently. It may be due to the absence
of proper numerical solver of the integro-differential equation of
resonant photon transfer in optical thick medium. The WENO solver
provided the first $x_{\pm}$-$\tau_0$ relation of DLA halo with
moderate and high optical depth.

\begin{figure}
\centering
\begin{center}
\includegraphics[width=8.0cm]{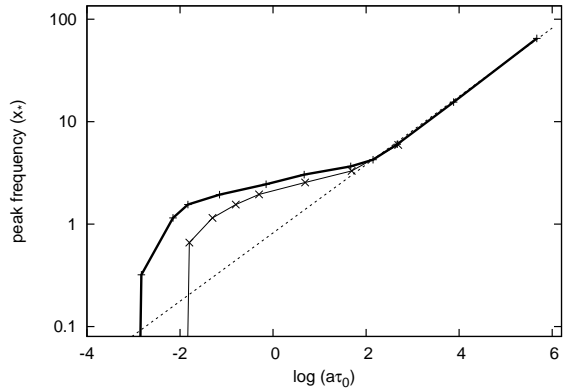}
\end{center}
\caption{The two-peak positions $x_{\pm}$ vs. $(a\tau_0)$ given by
MC simulation (data points) of a DLA halo with optical depth
$\tau_0$.  The parameter $a$ is taken to be  $5\times10^{-4}$(bold line) and
$5\times10^{-3}$ (light line). The dashed straight line of $\log
x_{\pm}$-$\log a\tau_0$ with slope 1/3 is to show the
$(a\tau_0)^{1/3}$-law.}
\end{figure}

The $x_{\pm}$-$\tau_0$ relation given by the MC simulation is
presented in Fig. 3, which shows that the $(a\tau)^{1/3}$-law is
significantly deviating from the MC results when $a\tau_0 \leq
10^2$, where the optical depth is smaller than $\tau_0=2\times10^4$ at $a=5\times10^{-3}$. This result
is the same as the WENO solution. In Fig. 3 the parameter range of [$a\tau_0]$
is larger than that of WENO solution [Fig. 4 of \citet{Roy10}].

In the range $10^{-2}<a\tau_0<10^2$, the $|x_{\pm}|$-$\tau_0$ relation
is almost flat with $|x_{\pm}|\simeq 2$. It is because the double-peaked
profile is from photons stored in the frequency range of $|x|<2$
and in the local thermal equilibrium state. The positions of the two peaks,
$x_{\pm }$, actually is about the same as the frequency range of the local
thermalization.  The frequency width $|x|\leq 2$ of the local thermal equilibrium
state is determined by the Doppler broadening, and very weakly dependent on
$\tau_0$. Thus, once the photons in local thermal equilibrium state are dominant,
we always have $x_{\pm}\simeq \pm 2$. This point can also be seen with Figs.
1 and 2, in which the positions of the two peaks are kept to be
$|x_{\pm}|\sim 2$ despite that the intensity of the flux increases
with time significantly.

When $a\tau_0<10^{-2}$, the curves of $|x_{\pm}|$ are no longer
determined by one variable $a\tau_0$, instead by variables $a$
and $\tau_0$, separately. The value of $|x_{\pm}|$ shows a quick drop
to zero at $a\tau_0 \sim 10^{-2}$ for $a=5\times 10^{-3}$, and $a\tau_0 \sim 10^{-3}$
for $a=5\times10^{-4}$. The W-F thermal equilibrium cannot be established in halos
with small optical depth $\tau_0<<10^2$, and therefore, photons from
these halos do not have double-peaked profile.

Since thermalization will erase all frequency features within the range $|x|\leq 2$,
the double-peaked structure doesn't retain information of the photon frequency
distribution within $|x|<2$ at the source. It is impossible to probe the
frequency profile for $|x|<2$ Ly$\alpha$ photons of the source from the
escaped Ly$\alpha$ photons. This property can also be used as a test of simulation code. 
That is, the simulation results should be independent of the profile of Ly$\alpha$ 
emission from the sources, only if the profile is non-zero within the range $|x|<2$, i.e. 
it should not matter whether the source is monochromatic, or has a finite width around 
$\nu_0$.

\section{Ly$\alpha$ photon emission: DLA model}

\subsection{Time dependent Ly$\alpha$ escape}

It is well known that the spatial transfer of Ly$\alpha$ photon in
optically thick halo is not simply a Brownian random walk. The time scale
of Ly$\alpha$ photon escape from optically thick halo is much shorter than
that of Brownian diffusion. It is because the spatial transfer depends on the
diffusion in frequency space. This is the so-called ``single longest excursion''
process \citep*{Adams72}. However, earlier estimates of the escaping
time scale based on ``single longest excursion'' can not describe the details of
the time-dependent behavior of photons escaping from optically thick medium.

If the central source of a DLA halo is assumed to be a photon flash, the
time dependence of the luminosity of the photon source is proportional to
$\delta(\eta)$.  Without scattering, the luminosity at the boundary of the halo
with size $r$  should still be a delta function as $\propto \delta(\eta-r)$,
i.e. it is also a flash, but with a retarded time $r$, which is the time needed
for a freely streaming photon from the center to the edge of the halo with speed
of light. Considering the effect of resonant scattering, the luminosity of escaped photons
will no longer be a flash. The light curve of the luminosity from such a source for a halo with size
$r=2\times 10^7$ and $ a=5\times 10^{-4}$ is shown in Fig. 4, in which $F(\eta)=\int f(\eta,x)dx$
is the flux integrated over all frequencies of escaped photons.

\begin{figure}
\centering
\begin{center}
\includegraphics[width=8.0cm]{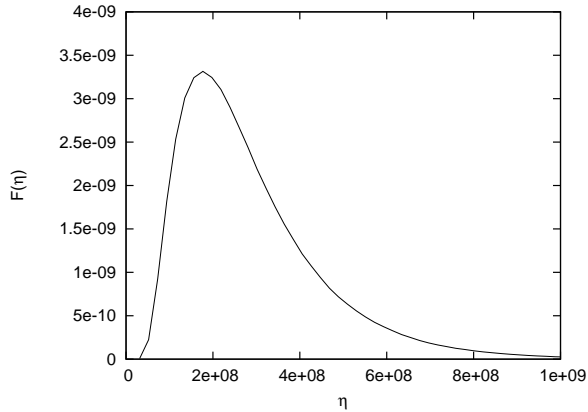}
\end{center}
\caption {The light curve of a flash of Ly$\alpha$ photons in a DLA
halo model of $ a=0.0005 $ with optical depth $\tau_0 =r_{\rm D}= 2\times 10^7$. One photon is
supposed to be emitted from a central source at time $\eta=0$, i.e. the integral
of the light curve is equal to 1. }
\end{figure}

Fig. 4 shows that the light curve lasts from time $\eta\sim 10^8$ to
$5\times 10^8$. The peak of light curve $F_{\rm max}$  is at
$\eta_{\rm max} \sim 2 \times  10^8$, and the time
duration $\Delta \eta$ of $F(\eta)>F_{\rm max}/2$, is also about $2\times 10^8$.
Since the source of photons does not contain any time scale, both the
numbers $\eta_{\rm max}$ and $\Delta \eta$ are from the size or optical
depth of the halo. The time scale $\eta_{\rm max}$ means that the retarded time
of the escaping of photons from the halo is as large as
$\sim 10 r$ or 10$\tau_0$. The amount $\Delta \eta$ means that the time-distribution
of the escaped photons are significantly spread out from a Delta
function $\delta(\eta)$ to $\Delta \eta \sim 10 r$.

Without resonant scattering, photons emitted from the source will escape from the halo
at time $\eta =r=\tau_0$. With resonant scattering, the majority of photons emitted from the
source will not escape from the halo until time $\eta=\eta_{\rm max}\sim 10^8$.
Therefore, a huge number of resonant photons is stored in the halo. The resonant nature of
Ly$\alpha$ photon scattering let the photons to stay in the halo with
a time scale equal to 10 times of the optical depth $\tau_0$ of the halo.

In our model the destruction processes of Ly$\alpha$ photons,
such as the two-photon process, are not considered, and
dust absorption is ignored too. The number of photons is
conserved. Thus, we have $\int F(\eta)d\eta=1$.
Therefore, the light curve, $F(\eta)$ of Fig. 4 can be understood as
the probability distribution of the time of Ly$\alpha$ photon escaped from a
$r=2\times 10^7$ halo. With dimensionless variables,
the curve of Fig. 4 is also the probability distribution of the total
length of the path of a photon transferring from $r=0$ to $r=2\times 10^7$.
In this context, $\eta_{\rm max}$ and $\Delta \eta$ can be used as the most probable
length of the path, and the variance of the distribution, respectively.
Considering that the most probable path length $\eta_{\rm max}$ and the
variance $\Delta \eta$ have the same order, the spatial transfer of Ly$\alpha$
photon in optically thick halo essentially is still a random process of diffusion.

The light curve of a flash source (Fig. 4) can easily be generalized to
arbitrary sources. Considering the total flux is linearly dependent
on sources, the total flux $L(\eta)$ is given by
\begin{equation}
L(\eta) =\int_0^{\eta} F (\eta - \eta' )s(\eta')d\eta'
\end{equation}
where $F(\eta)$ is the curve of Fig. 4, and $s(\eta)$ is the time-dependent
 Ly$\alpha$ photon flux of the sources.

\subsection{Escape coefficient}

For a source with stable luminosity of Ly$\alpha$ photons, escape
coefficient of Ly$\alpha$ photons emergent from a halo with size $r$
can be calculated by $f_{\rm esc}(\eta, r) \equiv F(\eta)/F_0 =\int f(\eta, r, x) dx$
if the luminosity of sources is normalized. Since the number of
Ly$\alpha$ photons is conserved,  the total number of escaped photons
should be equal to the total number of emitted photons by the source when a stable state is reached. Thus, the escape 
coefficient of a stable source should reach to 1
when $\eta$ is large enough.  However, before the system approaches
to stable state, the escape coefficient $f_{\rm esc}(\eta, r)$ can be
much less than one.

We have calculated the time dependent solution of the escape
coefficient $f_{\rm esc}(\eta)$ for a stable photon source with unlimited life time located
at the center of the halo with optical depth $\tau_0 = 2\times10^7$.
At time $\eta_{\rm max} = 2\times10^8$,
the escape coefficient $f_{\rm esc}(\eta) \sim 0.37$.
If we define the time scale $\eta_{\rm sat}$
of reaching saturated or stable state as
$f_{\rm esc}(\eta_{\rm sat}) = 0.95$, it yields $\eta_{\rm sat} = 6.1 \times 10^8$,
or $\sim 30 \tau_0$. Therefore,
$f_{\rm esc}$ is always significantly less than 1, when the age is less than about $30 \tau_0$ or 30
$r$.

\begin{figure}
\centering
\begin{center}
\includegraphics[width=8.0cm]{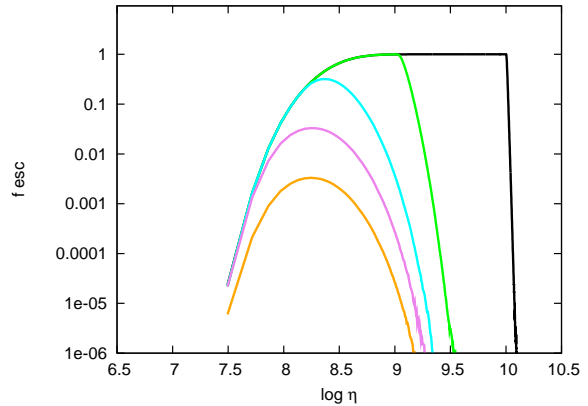}
\end{center}
\caption{Time dependent solutions of escaping coefficient $f_{\rm esc}(\eta)$
for photon sources located in the center of a DLA halo of $ T=10^4 $ K with optical depth
$\tau_0=2\times 10^7$. The life duration of the source $\eta_{\rm
life}$ is taken to be $\eta_{\rm life}=$ 10$^6$, 10$^7$, 10$^8$,
10$^9$ and 10$^{10}$, respectively, from bottom up using color code orange, violet,
cyan, green, and black. }
\end{figure}

Fig. 5 presents the time dependent solutions of escape coefficient
$f_{\rm esc}(\eta)$ for photon sources in a $T=10^4$ K neutral medium with limited life time to be $\eta_{\rm
life} = 10^6, 10^7, 10^8, 10^9$ and $10^{10}$, respectively, from
bottom up. It shows that the escape coefficient is always less than 1
if $\eta_{\rm life} < 10^7$. In this case, the time
scale of the escaping of Ly$\alpha$ photons in $\tau_0 =
2\times10^7$ halo is much larger than the life time of the source,
and therefore, photons emitted within the time duration $\eta=0$ to
$\eta=\eta_{\rm life}$ are fully spread over a time scale $\Delta
\eta\gg \eta_{\rm life}$. Thus, the escape coefficient of a source
with short life time is always much less than 1 when the DLA halo is
optically thick. This mechanism would be important for understanding
the observable features of high redshift objects such as GRBs or
first stars. Even when dust absorption is negligible, the escape
coefficient can be small, even very small, when the age of the
object is small.

It is interesting to see that although different curves of Fig. 5
correspond to very different $\eta_{\rm life}$, all the curves with
$\eta_{\rm life} < 10^7$ approach their maximum at about the same
time $\eta \sim 2 \times 10^8$. This time scale actually is the
$\eta_{\rm max}$ shown in Fig. 4. The stored photons yield a
delayed emission with time scale of about $\eta_{\rm max} \sim 2
\times 10^8$. The delay time is independent of the life-time
$\eta_{\rm life}$ of the source, but dependent only on the optical
depth of the halo.

\subsection{Surface brightness and  the size of DLA halo}

For a stable source, the time-independent surface brightness (SB) of a stable
source located in DLA halo is simply proportional to $r^{-2}_{\rm DLA}$,
where $r_{\rm DLA}$ is the radius of the halo. When the life time of
the source is short, the time-dependent behavior of the surface brightness, $SB(t)$,
is not simply proportional to $r^{-2}_{\rm DLA}$. To demonstrate this point,
we calculate the surface brightness of a source with life time 10$^6$ years and
Ly$\alpha$ luminosity $L=10^{42}$ erg s$^{-1}$  in a medium of $a=0.0005$. The results are
presented in Fig. 6, in which the size of the halos is taken to be
$r_{\rm DLA} = 1 $kpc, 10 kpc, 100 kpc, and 1 Mpc. In Fig. 6, the optical depth of
the DLA halo surrounding the source is assumed to be $\tau_0= 2\times 10^7$, which
correspondes to a typical DLA halo with column density
$N_{\rm HI}= 2 \times 10^{20}$ cm$^{-2}$ (Eq.(1)). 

\begin{figure}
\centering
\begin{center}
\includegraphics[width=8.0cm]{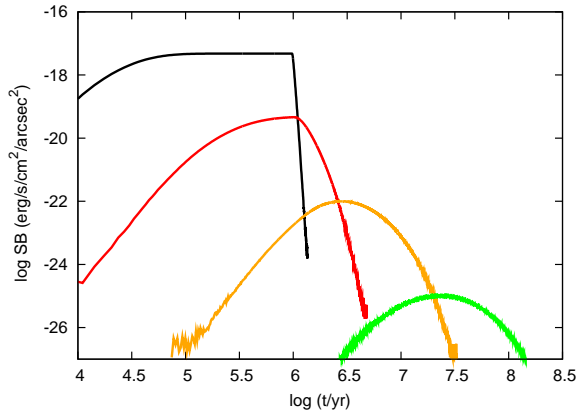}
\end{center}
\caption{Time dependent surface brightness of a central
source surrounded by a DLA halo of $ T=10^4$K with optical depth $\tau_0=2\times10^7$.
The source emits  Ly$\alpha$ photons with  stable flux $L=10^{42}$ erg s$^{-1}$
in the time range from $t= 0$ to $t=10^6$ years. The size of DLA halo
is taken to be $r_{\rm DLA} = 1 $kpc, 10 kpc, 100 kpc and 1 Mpc, respectively
from top down using color codes black, red, orange and green.
}
\end{figure}

From Fig. 6, we first see that the curves do not show the behavior of being
proportional to $r^{-2}_{\rm DLA}$. The maximum surface brightness of halo with
$r_{\rm DLA}=$ 1 kpc is about $10^{-17}$ erg s$^{-1}$ cm$^{-2}$ arcsec$^{-2}$, while
the maximum surface brightness of $r_{\rm DLA}=$ 1 Mpc is only
$\sim 10^{-25}$ erg s$^{-1}$ cm$^{-2}$ arcsec$^{-2}$. That is, when $r^{-2}_{\rm DLA}$
decreases by a factor $10^{6}$, the maximum surface brightness descreases by a factor
10$^8$.

Secondly, the shapes of the curves of Fig. 6 for different $r_{\rm DLA}$ are very
different. The curve of $r_{\rm DLA}=$1 kpc shows saturation at $t> 3 \times10^4$ years, while
all others don't have a saturated phase.

These time-dependent behaviors are also due to the time scale of Ly$\alpha$ photon
transfer. For a halo with given column density $N_{\rm HI}$, or $\tau_0$, the time
scale of the Ly$\alpha$ photon transfer is about
$\eta_{\rm max} \sim 10\tau_0$, which is independent of the size $r_{\rm DLA}$ of the halo.
However, $\eta_{\rm max} \sim 10\tau_0$ corresponds to a physics time
$t_{\rm max}=\eta_{\rm max} r_{\rm DLA}/c$, which is $r_{\rm DLA}$-dependent.
For DLA halo with $r_{\rm DLA}=$1 kpc, the time scale
$t_{\rm max}=\eta_{\rm max} r_{\rm DLA}/c\sim 3\times10^4$ years. It is much less than 10$^6$
years, and therefore, the surface brightness is saturated when $t> 3\times10^4$ year.
For DLA halo with $r_{\rm DLA}=$100 kpc, the time scale
$t_{\rm max}=\eta_{\rm max} r_{\rm DLA}/c\sim 3\times10^6$ years. It is larger than
the life-time of the source, and therefore, the surface brightness can not approach a
saturated state. For DLA halo with $r_{\rm DLA}>$ 100 kpc, the time scale
$t_{\rm max}=\eta_{\rm max} r_{\rm DLA}/c$ would be much larger than 10$^6$ years. The source is more like
a single flash and its radiative transfer is highly time dependent (Fig. 4).

Thus, we may conclude that for DLA halos with a given column density, the surface brightness
will be proportional to $r^{-\kappa}_{\rm DLA}$ and $\kappa>2$. This result would be useful to estimate
the size of the DLA halo with observed surface brightness and the model of the sources.

\section{Ly$\alpha$ photon emission: IGM model}

\subsection{Ly$\alpha$ escape}

The mechanism of the escaping of Ly$\alpha$ photons from expanding
opaque IGM at high redshift is different from that of DLA halos.
For the former, besides the diffusion in frequency space caused by the the resonant
scattering, we should also consider the frequency redshift caused by cosmic
expansion. Photons will escape from Gunn-Peterson trough, once their
frequency is redshifted enough.

\begin{figure}
\centering
\begin{center}
\includegraphics[width=8.0cm]{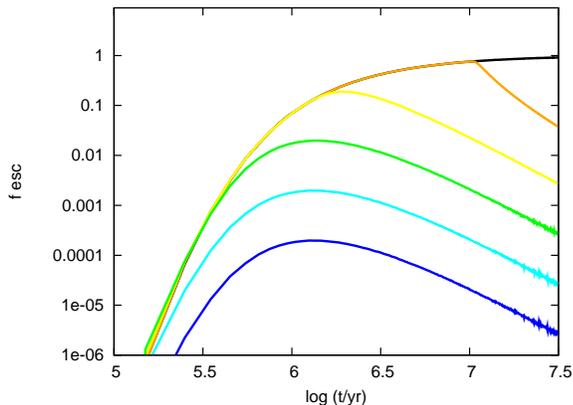}
\end{center}
\vspace{-5mm} \caption{Time dependent solutions of escaping coefficient
$f_{\rm esc}(t)$ for a source at redshift $1+z=10$ surrounded by the IGM of $T=100$K
of the expanding universe. The time duration of the emission of the source
is taken to be $t_{\rm life}=10^3$, 10$^4$, 10$^5$, 10$^6$, 10$^7$, and 10$^8$
years, respectively from bottom up using color codes blue, cyan,green, yellow,
orange, and black. $t$ is the delayed time of an escaped Ly$\alpha$ photon
 with respect to the first escaped free streaming photon from the source, measured at
 the redshift of the source.}
\end{figure}

Similar to the previous section, we calculate the escape coefficient
using the number of escaped photons. Fig. 7 presents the
time-dependent solutions of the escape coefficient $f_{\rm esc}(t)$
for sources at redshift $1 + z = 10$ surrounded by the IGM of Hubble
expanding universe. The number density of neutral hydrogen is given
by the cosmology parameters of the $\Lambda$CDM model. The IGM is
assumed to be not reionized yet. The time duration $t_{\rm life}$ of
photon source has been explored with values $t_{\rm life} =
10^3 , 10^4 ,10^5 , 10^6 , 10^7$, and $10^8$ years. For sources at
$1 + z = 10$, $t_{\rm life}$ cannot be larger than the age of the
universe $\sim 10^8$ years. The source is starting to emit photons at $t=0$.

The curves of Fig. 7 look very similar to those of Fig. 5. Therefore, one
can explain Fig.7 in the same way as Fig. 5, replacing the
optical depth of the DLA halo with the Gunn-Peterson optical depth (Eq.(A5)) of the
expanding IGM at $(1+z)=10$. In Fig. 7, all the short life time curves with
$t_{\rm life} < 10^7$ years reach their
maximum at about the same time $t_{\rm max} \sim  10^6$ years, which is the
time scale of Ly$\alpha$ photon escape from the Gunn-Peterson trough in
an expanding IGM at $1+z=10$. One can then conclude that the escape
coefficient of sources with life time less than $t_{\rm max} \sim 10^6$ years
should always be less than 1, even without dust absorption.

The curve of the shortest life time ($t_{\rm life}=10^3$) of Fig. 7
can be thought to represent the light curve of a flash source in IGM model, like that of
Fig. 4 for DLA model. The curve of
$t_{\rm life}=10^3$ of Fig. 7 has a long tail. The long tail basically is a
power law of $t$,  and is a joint result of radiative transfer and Hubble expansion velocity field.
 As mentioned in \S
4.1, for the light curve of Fig. 4, the maximum $\eta_{\rm max}$
and variance  (or the width of the light curve) $\Delta\eta$ are about the
same. Yet, the power law long tail leads to the maximum $t_{\rm max}$ of
the curve of $t_{\rm life}=10^3$ of Fig. 7 to be much less than the width
of the curve. Since the suppression of escape coefficient is
mainly given by the width of the light curve. Therefore, the power law long
tail of Fig. 7 implies that the stable state of $f_{\rm esc}=1$ takes
a much longer time to approach, or can never be approached within the age of the
universe.

Although Fig. 7 shows that the escape coefficient of sources with
$t_{\rm life} \geq  10^6$ years is larger than those of sources with
$t_{\rm life} \leq  10^6$, it does not mean that the former is
easier to be observed than the later. This point can more clearly be
revealed with the time-dependent solution of surface brightness.

\subsection{Surface brightness}

We calculate the mean surface brightness
 $SB(t)\equiv \frac{L}{\pi <r_{\rm esc}^2>}$ defined as the flux divided
by the averaged area of the source, where $r_{\rm esc}$ is 
the projected distance to the source on
the sky at which a photon escapes. Thus, $\pi <r_{\rm esc}^2>$(t) is the
mean of the area at time $t$ on the plane perpendicular to the line of
sight.  The result is presented in Fig. 8, which plots the $SB(t)$ for
sources with Ly$\alpha$ luminosity (erg/sec) and life time (year) paired as:
($10^{46}$, 10$^3$), ($10^{45}$, 10$^4$), ($10^{44}$, 10$^5$),
($10^{43}$, 10$^6$), ($10^{42}$, 10$^7$), and ($10^{41}$, 10$^8$).
That is, all sources emit the same amount of  $2\times10^{67}$ Ly$\alpha$ photons in their
whole life span. The sources start to emit at time $t=0$.

\begin{figure}
\centering
\begin{center}
\includegraphics[width=8.0cm]{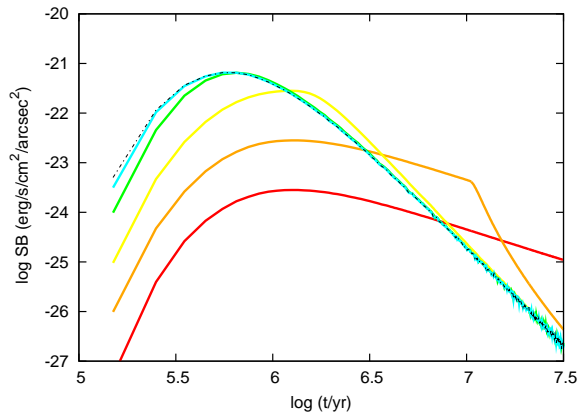}
\end{center}
\caption{
Time dependent surface brightness from a central
source surrounded by IGM of $T=100$ K in expanding universe. The source
is assumed to be located at redshift $1 + z = 10$, and emitting a
total of $2 \times 10^{67}$ Ly$\alpha$ photons from time $t = 0$. The time duration
of the emission of the source is taken to be $t_{\rm life} = 10^4, 10^5, 10^6, 
10^7$, and $10^8$ years, respectively from top down using color codes
cyan, green, yellow, orange, and red. The flash source ($t_{\rm life} = 10^3$ years) is drawn
with black dashed line.}
\end{figure}

Fig. 8 has two remarkable features. First, the four curves of
($10^{46}$, 10$^3$), ($10^{45}$, 10$^4$), ($10^{44}$, 10$^5$), and
($10^{43}$, 10$^6$) are almost the same. The common curve has a maximum at
$t \sim 10^6$ year and then starts decaying with a power law $SB(t)\propto t^{-4}$. The
time scales of the maximum and the power law tail are about the same as that
of the light curve of
Fig. 7 with $t_{\rm life}=10^3$. Therefore, for sources with life time less than
$t \sim 10^6$ years, the surface brightness is only dependent on the
total number $N$ of Ly$\alpha$ photons emitted from the source, regardless of their
life time. This is because the photons should wait for about
$t\sim 10^6$ years before their escaping from the IGM. The stored photons are locally
thermalized, and therefore, the information of the ``age'' of the photons emitted at
time $< 10^6$ will be forgotten during the thermolization. This property actually is also
valid for the sources of ($10^{42}$, 10$^7$), and ($10^{41}$, 10$^8$). For these two cases,
the total numbers of Ly$\alpha$ photon emitted in $t \leq  10^6 $ years are smaller,
respectively are $2\times10^{66}$, and $2\times10^{65}$ and therefore, their $SB(t)$ at
$t \sim 10^6 $ is less than that of $N=10^{67}$ by factors 10 and 100,
respectively in the figure. 
The curves of $t_{life}=10^4, 10^5$ years  become scalable from  $\sim 1$ Myr.
The $t_{life}=10^6$ yr curve is just starting to move away from the flash source
solutions. The relaxation time looks to be around $0.3-1$Myr when the flash source reaches its
maximum. This is in agreement 
with the estimate from \citet{Rybicki94} where the relaxation time is
$(\frac{a}{\gamma^4})^{\frac{1}{3}}*t_s \sim 0.3$ Myr 
for our chosen redshift ($1+z=10$) and temperature ($T=100 K$).
Fig. 8 shows various scaling relations. For example, for the flash source of $t_{life}=10^3$yr, the asymptotic slope sets
in from $\sim 1$ Myr as a result of radiative transfer in Hubble expansion velocity field.
For the $t_{life}=10^8$ yr curve in the figure, another straight part sets in from 1 Myr 
to 50 Myr even before the asymptotic slope is reached,
as a joint result of photon emission, photon transfer and Hubble expansion. These time scales are
comparable to the
possible life time of the sources, which was 
previously pointed out by  \citet{Higgins09}.

The second feature of Fig. 8 is the monotonous decrease of $SB(t)$
when $t$ is large, regardless of the life time of the source. It is very
different from all the curves of Fig. 5, which will
approach a stable or saturated state if the life time of the source is long enough.
For any sources in expanding neutral IGM, the
mean surface brightness $SB(t)$ does not approach a saturated or
stable state. In other words, a time-independent solution of the
surface brightness doesn't exist in the IGM model. This is simply
because the increase of volume of the expanding IGM, which stores
Ly$\alpha$ photons, is faster than the number of Ly$\alpha$ photons
redshifted to frequency $x < -2$. This feature is similar to the
evolution of ionized halos around a UV photon source in expanding
universe. The ionized radius can never approach a stable state
required by a Str\"omgren sphere, because the increase of the
ionized radius is always lower than the comoving
velocity\citep*{Shapiro87}. Therefore, in Fig. 8 there is no flat section for
every curve.
The maximum of the surface
brightness of a Ly$\alpha$ photon source with luminosity $L$ at
$1+z=10$ scales approximately as $\sim 10^{-21}\times(L/10^{43}{\rm erg/s})$ \ erg
s$^{-1}$ cm$^{-2}$ arcsec. We should emphasize that the maximum can
be reached only for sources with age equal to about 10$^6$ years. The
surface brightness would be less than the maximum when the source
age is younger or older than 10$^6$ years. That is, in term
of surface brightness, a stable source will yield a time-dependent
curve.

\section{Ly$\alpha$ photon absorption}

\subsection{Time dependent red damping wing: DLA model}

A typical problem of Ly$\alpha$ absorption at high redshift is the red damping wing, or
HI damping wing of high redshift sources. The optical afterglow of GRBs at high redshifts
has shown this feature(e.g. \citealt{Totani06,Salvaterra09}).
We calculate the time-dependence of the red damping wing of a DLA halo with radius
and optical depth $\tau_0$ to be $r=\tau_0=2\times10^7$.  The frequency spectrum
of the central photon source is assumed to be flat with flux equal to one. If
the source starts to emit photons at time $\eta=0$, photons without undergoing
scattering or collisions will escape from the halo at time $\eta_c=2\times 10^7$.
The profiles of red damping wing at times later than $\eta_c$ are plotted in Fig. 9,
in which we take $\eta=(1+ y)\eta_c$, where $y=0.01$, 0.5, 2, 5, 10 and 50.

\begin{figure}
\centering
\begin{center}
\includegraphics[width=8.5cm]{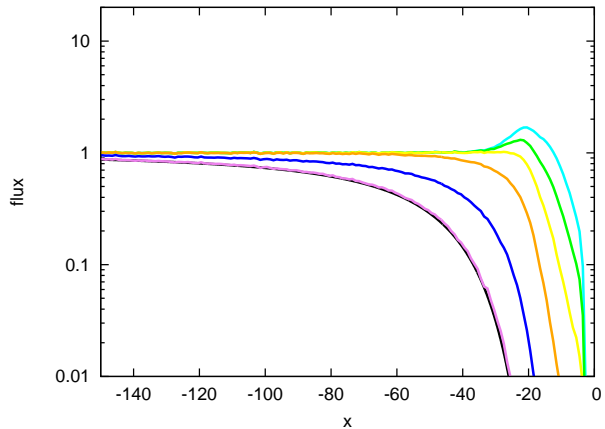}
\end{center}
\caption{Time dependent red damping wings formed in a DLA halo of parameter $ a=0.0005 $ with size $r$ and
optical depth $\tau_0$ set to be $r= \tau_0=2\times10^7$. The central source
starts to emit photons with a flat spectrum at $\eta=0$. The profiles
of the red damping wing are shown at times $\eta=(1+ y)\eta_c$,
where $\eta_c=r=\tau_0$ is the time of free flight from $r=0$ to
$r$, and  $y$ is taken to be $0.01$, 0.5, 2, 5, 10, and 50, respectively
from bottom up using color code violet, blue, orange, yellow, green, and cyan.
The black curve, which is almost the same as the violet curve, is given by the
Voigt absorption, i.e. $f(x)=e^{-\tau(x)}$ and $\tau(x)$ is from Eq.(1).}
\end{figure}

Fig. 9 shows that the red damping wing at time $\eta \leq 1.01 \eta_c$ can be well
described by a Voigt profile $f(x)=e^{-\tau(x)}$, where $\tau(x)$ is given by Eq.(1).
Therefore, it would be fully reasonable to fit the red damping wing of GRB's optical
afterglow with a Voigt absorption profile, because the red damping wing is measured
only a few hours or a few days after the GRB explosion, the time $\eta$ is very close to
$\eta_c$. Even when $\eta = 1.5 \eta_c$ (the blue line in Fig.9), the
red damping wing can still be approximately fitted by a Voigt profile. However, the
fitting at time $\eta = 1.5 \eta_c$ will yield a smaller $\tau_0$ than that of the fitting
at $\eta =1.01 \eta_c$. Therefore, the column density of HI atoms given by the fitting
at $\eta = 1.5 \eta_c$ is underestimated.

The red damping wing at $\eta\geq 5 \eta_c$ shows a shoulder with
frequency similar to that given by Fig. 3 with parameters
$a=5\times10^{-4}$ and $\tau_0=2\times 10^7$. Therefore, the shoulder is
from the stored resonant photons. Fig. 9 shows that the curve of
red damping wing has become saturated at time $\eta \geq 50\eta_c$.
This is larger than the time scale $\sim 10\eta_c$ shown in Fig.
4. It is because the photons with frequency $|x|> 2$ of the
continuous spectrum can also be stored. According to the
redistribution function (A6), the probability of transferring a
$|x|>2$ photon to $|x|<2$ is larger than that from $|x|<2$ to
$|x|>2$, and therefore, in frequency space the net effect of the
resonant scattering is to bring photons of the continuous spectrum
background to become Ly$\alpha$. More photons be stored leads to the
larger time scale of the saturation.

\subsection{Time dependence of red damping wing: IGM model}

The problem of the red damping wing for the IGM model is very
different from that of the DLA model. If the frequency spectrum of
source is flat, cosmic redshift will continuously provide Ly$\alpha$
photons from blue side of the spectrum. Since the time scale of the
W-F thermalization is much shorter than the time scale of the cosmic
expansion, the photons moved into the frequency range $\sim \nu_0$
from $> \nu_0$ is quickly thermalized. Thus, a huge number of
thermalized photons is stored in the frequency range $|x|<2$
\citep{Roy09b,Roy10}. This process marks the difference between the
IGM and the DLA models, and also the difference between our IGM model and IGM models of some others
\citep{Loeb99}, in
which the source emits only $\nu_0$ photons.

\begin{figure}
\centering
\begin{center}
\includegraphics[width=7.8cm]{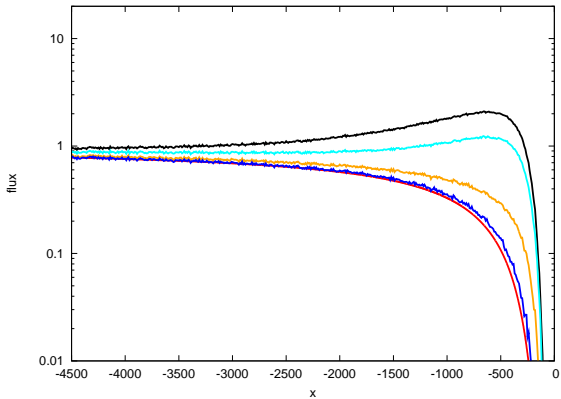}
\end{center}
\caption{The time dependence of the red damping wing given by simulations in an
expanding IGM model. The central source  at $ 1+z=10$ starts to emit photons with flat spectrum at
$t=0$ into IGM of T=100K. The curves correspond to $t=10^5$, 10$^6$, $5\times10^6$, and 10$^7$ years, respectively
from bottom up using color codes blue, orange, cyan, and black. The red line is given 
by pure absorption $f(x)=e^{-\tau(x)}$ where $\tau(x)$ is from Eq.(C1).
}
\end{figure}

Fig. 10 shows the red damping wings of a source in T=100 K IGM at $1+z=10$ with
age $t=0, 10^5$, 10$^6$, $5\times10^6$, and 10$^7$ years. When the time is
less than $10^5$ year, the Ly$\alpha$ photons have not been
significantly stored yet, and the profile of red damping wing is
about the same as a pure absorption, which has been addressed  by
\citet{Mir98}. Once the time is larger than $10^6$ year, the stored
photons yield a shoulder at peak frequency $-x \sim 500$.
Fig. 10 also shows that the profile of the red damping wing seems
not yet to approach a saturated or time-independent state at
$t=10^7$ years.

\section{Discussions and conclusions}

The time-dependent behaviors in physics and frequency spaces of the
Ly$\alpha$ photons emergent from optically thick medium have been
extensively studied with Monte Carlo method. The first conclusion is
that time-dependent solutions are essential not only for strong time
dependent sources like GRBs, but also for stable sources, especially
when the cosmic expansion needs to be considered. Cosmic expansion
makes the radiation transfer equation not invariant with respect
to the transformation of time shift. Consequently, the radiative transfer
equation doesn't have time-independent solutions in principle. Time-dependent
behavior are essential.

With the model of IGM in the Hubble expanding universe, we show
that the time-dependence of the brightness of a stable source at high
redshift is like that of the light curve of an "explosion", i.e. in the first
phase, the surface brightness is increasing and approaching a
maximum, and then, in the second phase, it decays monotonously.
Although our calculation is only based on a model source at redshift
$(1+z)=10$, we believe that the feature of the nonexistence of
a time-independent solution, or the nonexistence of a saturated state of the surface
brightness by IGM scattering, would be hold for various high redshift sources.

The IGM model may be too simple and ideal. Many effects have not
been considered, such as the effects given by the inhomogeneity of
the HI density distribution, the nonuniform ionization, the bulk
velocity, the wind and the turbulence of the IGM field, and the observational
aperture, etc. However, one point seems to be clear that under these
circumstances, stable solutions will not exist. All these
effects should be studied with time-dependent solutions of the
Ly$\alpha$ photon transfer. Since some high redshift Ly$\alpha$
emitters have already been observed, static solutions may not be
enough to understand these observations.

For the DLA model, the saturated state or the time-dependent state, does
exist. Therefore, it is reasonable to find this state directly by
time-independent solution. However, time-independent solutions
can not tell us the suppression of the escape coefficient in the time
range of $\eta < 10 \tau_0$, $\tau_0$ being the optical depth of the
halo. When $\tau_0$ is large, the escape coefficient will be
substantially suppressed for time range comparable to the age of the
universe. This is equal to say that the escape coefficient is much
less than one in all time.

Low escape coefficient is often explained by dust
absorption. Dust extinction is effective only when the medium is
optically thick. Therefore, once we use the model of dust absorption
at high redshift, the time-dependent behavior must be considered. A
WENO time-dependent solution of the Ly$\alpha$ photon transfer in
dusty medium will be reported soon.

We didn't consider the Ly$\alpha$ photons from the recombination in
the ionized halo around the center source. Simply changing the
luminosity of the central light source by adding the photons from
stable recombination in the halo is a poor approximation, as the
time scale of the recombination is not small. Therefore, with the
DLA model, time-dependent solutions are also essential for
understanding Ly$\alpha$-related phenomenon.

\section*{Acknowledgments}

This work is supported by the Ministry of Science and Technology of
China, under Grant No. 2009CB824900.

\appendix

\section{Radiative transfer of Ly$\alpha$ photons}

 With dimensionless variables, the specific intensity $I$ of photon number
is a function of $\eta$, $r$, $x$ and $\mu$.
When the optical depth is large, we can take the Eddington approximation.
Radiative transfer equations are \citep*{Roy09c}
\begin{eqnarray*}
{\partial j\over\partial \eta} + \frac{\partial f} {\partial r} & =
& - \phi(x;a)j + \int \mathcal{R}(x,x';a)j dx'+  \gamma
\frac{\partial
j}{\partial x}+ r^2S,\\
\end{eqnarray*}
\begin{equation}
\end{equation}
\begin{equation}
\frac{\partial f}{\partial \eta} + \frac{1}{3}
\frac{\partial j} {\partial r} - \frac{2}{3}\frac{j}{r} -\gamma \frac{\partial f}{\partial x} =  -
\phi(x;a)f.
\end{equation}
where  $j,f$ are the rescaled quantities of physical intensity, $j(\eta, r,x)= r^2 j_p =r^2\frac{1}{2}\int_{-1}^{+1}I(\eta,r,x,\mu)d\mu$ is
the angularly averaged specific intensity and
$f(\eta, r,x)= r^2 f_p=r^2\frac{1}{2}\int_{-1}^{+1}\mu I(\eta,r,x,\mu)d\mu$ is flux. 
The mean intensity $j(\eta,r,x)$ describes the $x-$photons
trapped in the halo by the resonant scattering, while the flux
$f(\eta,r,x)$ describes the photons in transit. The parameter $\gamma=1/\tau_{GP}$
can be calculated as
\begin{equation}
\tau_{GP} = 4.9 \times 10^5 h^{-1} f_{HI} \left ( \frac{0.25}{\Omega_M}
\right) ^{\frac{1}{2}} \left(\frac{\Omega_b h^2}{0.022}\right)
\left(\frac{1+z}{10}\right)^{\frac{3}{2}}.
\end{equation}
The term $S$ describes a constant physical photon source.
Since Eqs.(A1) and (A2) are linear
with respect to $j$ and $f$, the solutions of $j$ and $f$ for sources
$S(\eta, x)$ can be given by a superposition of solutions $j(\eta, r, x; \eta_0, x_0)$ and
$f(\eta, r, x; \eta_0, x_0)$, which are the solutions of the source
$S=\delta(\eta-\eta_0)\delta(x-x_0)$.

The resonant scattering is described by the redistribution function
$\mathcal{R}(x,x';a)$ which is the probability of a photon absorbed at the
frequency $x'$, and re-emitted at the frequency $x$. It depends on the
details of the scattering \citep*{Henyey41, Hummer62, Hummer69}. If we
consider coherent scattering without recoil, the redistribution function
with the Voigt profile  is
\begin{eqnarray}
\lefteqn{ \mathcal{R}(x,x';a)=\frac{1}{\pi^{3/2}}\int^{\infty}_{|x-x'|/2}e^{-u^2} } \\ \nonumber
 &  &
\left [\tan^{-1}\left(\frac{x_{\min}+u}{a}\right)-\tan^{-1}\left(\frac{x_{\max}-u}{a}\right
)\right ]du
\end{eqnarray}
where $x_{\min}=\min(x, x')$ and $x_{\max}=\max(x,x')$. This re-distribution function
 is normalized as $\int_{-\infty}^{\infty} \mathcal{R}(x,x')dx'=\phi(x,0)
=\pi^{-1/2}e^{-x^2}$.

Thus, the frequency of photons will be changed during the evolution
governed by Eqs.(A1) and (A2), while the total number of photons is conserved. That is,
the destruction processes of Ly$\alpha$ photons, such as the two-photon process
\citep*{Spitzer51, Osterbrock62} and dust absorption,
are ignored in equations (A1),(A2).

\section{Monte Carlo Solver}

Monte Carlo simulation starts with releasing a photon at the source placed along  a
random but isotropic direction. The frequency distribution
of the new photon follows that of the source, either a Gaussian distribution with a
Doppler core, or a continuum.
Once the photon enters the gas medium, the length of free path is determined by
calculating the optical depth variable
 $\tau$ traveled during the free flight,
following the distribution function $e^{-\tau}$. It is then straight forward to
convert from $\tau$ to distance.
The location of the next scattering is thus determined. If the place is
outside the HI cloud in a halo model, or if the traveled optical depth is larger
than Gunn-Peterson optical depth in an expanding IGM model, the photon is labeled escaped.

At the site of scattering, the velocity of the HI atom is generated by two steps.
First, the velocity components
$v_x$ and $v_y$ (normalized to Doppler velocity $V_D$, and $z$ is the propagation
direction of the photon) are generated following a Maxwellian distribution $e^{-{v_x^2}}$.
Second, the velocity $v_z$ is generated following the distribution:
$
f(v_z) \propto \frac{e^{v_z^2}}{(x-v_z)^2+a^2}
$
which is the joint requirement of Gaussian distribution and Lorentz profile for the rest
frame cross section of resonant scattering.
The direction of the resonantly scattered photons is assumed to be isotropic, but
can be easily adapted to other types of angular dependence.
Once the direction is generated, frequency of the outgoing photon can be calculated.
Using the notation of $x$, $x'$ to represent the laboratory frequency of the incoming
and outgoing photon, we have
$x' = x-v{\rm cos}\eta + v {\rm cos}\eta {\rm cos}\theta +
v {\rm sin} \theta {\rm sin} \eta {\rm cos}\phi - b(1-\rm{cos}\theta)$
where $b=\frac{h\nu_0}{m_e V_{D} c}$ is the recoil parameter, $\eta$ is the angle
between the incoming and outgoing photons,
$\theta$ and $\phi$ are the two spherical coordinates of the outgoing photon where
 the coordinates are chosen such that the incoming photon is in $z$ direction.
(We follow Field 1959's scattering geometry and notations.)
With this new set of frequency and direction of the photon, we repeat the above
procedures of calculating the next scattering and the determination on escape.
Each photon is followed all the way along its path until it escapes.

Since the effectiveness of generating $v_z$ determines crucially the speed of calculation,
special algorithms have been proposed (\citet{Zheng02}, ZM02 hereafter).
We basically follow ZM02's algorithm for medium to large $x$ ($0.6 \leq x \leq 17$).
For smaller $x$ ($x<0.6$), methods of plain rejection (not employing ZM02's algorithm)
is used.  For very large $x$ ($x > 17$),  our treatment for $u>u_0$ is similar to ZM02,
but for $u\leq u_0$, we switch the roles of the two functions,
using the distribution function $e^{-v_z^2}$ as the transformation method
to generate $v_z$, and then use $\frac{1}{(x-v_z)^2+a^2}$ as the comparison function to reject.

Twenty million of photons are experimented by Monte Carlo simulation for each model.
Simulations are performed only for sources with single flash of photons. In these
simulations, a time stamp can be recorded for each photon at each step of collision,
and at its escape.  For sources of arbitrary time dependence, a new random variable
is used representing the birth time of the photon.
The time stamps can be generated by adding a photon's birth time to
the recorded time stamps from a single flash simulation.
 Furthermore, with, say, $10^4$ trials of randomly generated birth time 
for each original photon in a single-flash simulation,
 we form a subgroup of $10^4$ photons, which have the same
history of collisions but happen at different epochs.
By this way we greatly improve the very low usage rate
when the data of each recorded photon is coupled with only one birth time.
Statistics over this enlarged group of photons give better continuity
 and smaller Poisson errors on the surveyed quantities, but do not add new physics.

\section{Gunn-Peterson optical depth}

The free path of a Ly$\alpha$ photon in a Hubble streaming IGM can be derived by using
GP optical depth for frequency $x$ measured at source redshift $z$, which is
\begin{equation}
\tau(x,z)=\tau_{GP}(z) \int_{-\infty}^x \phi(x) dx
\end{equation}
where the Gunn-Peterson optical depth parameter is given by equation (A3).

In MC simulations, the Hubble streaming always makes frequency $x$ lower and GP
optical depth smaller.
 Suppose the photon frequency is $x_A$ immediately after the last scattering and $x_B$
before the next scattering,
 the change of GP optical
 depth $\delta \tau$ during the free flight is always negative and equals to -1 on average.
The particular value of each $\delta \tau$ is determined by generating a random number
$\delta \tau \le 0$ following the distribution function $e^{\delta \tau}$. With this
$\delta \tau$ we can derive
the new frequency $x_B$ by solving the equation
\begin{equation}
\delta \tau=\tau_{GP}(z) \int_{x_A}^{x_B} \phi(x) dx
\end{equation}
By compiling a large data table of $\tau(x,z)$ and using linear interpolation,
$x_B$ can be solved fast.

The free length $l$ before the next scattering can be found by considering
that the frequency change is caused
by Doppler effect of Hubble motion,
\begin{equation}
x_B-x_A= - \frac{H(z) l}{c}\frac{\nu_0}{\Delta \nu_D}
\end{equation}
where Hubble constant at redshift $z$ is
\begin{eqnarray}
\lefteqn {
H(z)=H_0\sqrt{\Omega_M(1+z)^3+\Omega_R(1+z)^2+\Omega_{\Lambda}} }\\
 \nonumber
 & & \approx H_0\sqrt{\Omega_M}(1+z)^{\frac{3}{2}}
\end{eqnarray}
in which all $\Omega_M, \Omega_R, \Omega_{\Lambda}$ refer to values at present.

\end{document}